\documentstyle[prl,aps,preprint]{revtex}
\begin{document}
\tighten

\title{IMPLICATIONS OF COLOR GAUGE SYMMETRY 
\\ FOR NUCLEON SPIN STRUCTURE}
\author{Pervez Hoodbhoy, Xiangdong Ji and Wei Lu}
\bigskip

\address{
Department of Physics \\
University of Maryland \\
College Park, Maryland 20742 \\
{~}}

\date{UMD PP\#99-012 ~~~DOE/ER/40762-158~~~ August 1998}

\maketitle

\begin{abstract}
We study the chromodynamical gauge symmetry 
in relation to the internal spin structure
of the nucleon.  We show that 1) even in the helicity eigenstates  
the gauge-dependent spin and orbital angular momentum 
operators do not have gauge-independent matrix 
element; 2) the evolution equations for the gluon 
spin take very different forms in the  Feynman 
and axial gauges, but yield the same leading behavior 
in the asymptotic limit; 3) the complete 
evolution of the gauge-dependent orbital 
angular momenta appears intractable 
in the light-cone gauge. We define a new gluon 
orbital angular momentum distribution $L_g(x)$ which 
{\it is} an experimental observable and has a simple
scale evolution. However, its physical 
interpretation makes sense only in the light-cone
gauge just like the gluon helicity
distribution $\Delta g(x)$. 

\end{abstract}
\pacs{xxxxxx}

\narrowtext
 
The spin structure of the nucleon has been 
a subject of intense debate for about ten years.
Much progress has been made in both experimental
and theoretical frontiers \cite{review}.  
However, some of the fundamental theoretical 
issues remain unsettled, as exemplified by
a number of recent works in the literature.
In particular, in analyzing the spin structure of the nucleon, 
color gauge invariance
is still the cause of some confusion. 
In this paper, we intend to explore several
important and relevant issues in detail.

To set the stage, let us briefly recall
the forms of angular momentum operator
in quantum chromodynamics (QCD). In Ref. \cite{ji1}, a natural 
gauge-invariant expression is introduced
\begin{equation}
    \vec{J}_{\rm QCD} = \vec{J}_{q} + \vec{J}_g \ ,
\end{equation}
where
\begin{eqnarray}
     \vec{J}_q &=& \int d^3x ~\left[ \psi^\dagger
     {\vec{\Sigma}\over 2}\psi + \psi^\dagger \vec{x}\times
          i\vec{D}\psi\right]
     \ ,  \nonumber \\
     \vec{J}_g &=& \int d^3x ~\vec{x} \times (\vec{E} \times \vec{B}) \ .
\label{ang}
\end{eqnarray}      
The quark contribution $\vec{J}_q$ contains two terms. 
The first term is obviously the quark spin as 
$\vec{\Sigma}={\rm diag}(\vec{\sigma},{\vec\sigma})$ is the four-dimensional
generalization of the familiar Pauli spin matrices. The second term 
is the quark {\it kinetic} 
(or mechanical) orbital angular momentum, in which 
the covariant derivative $\vec{D} =- \vec{\nabla} + ig
\vec{A}$ originates from the quark kinetic momentum \cite{fey}. 
We recall that in a gauge  theory 
the kinetic momentum appears to be more physical than the dynamical 
(or canonical) one, the latter corresponding  to a partial 
derivative $-i\vec{\nabla}$ in quantum mechanics\cite{fey}. 
The gluon contribution to the angular momentum 
$\vec{J}_g$ contains the well-known Poynting vector,
$\vec{E}\times \vec{B}$, the momentum density of
the radiation field. Some recent studies 
in terms of the above form of angular momentum 
operators can be found in Refs. \cite{ji2,bro,sin}.

The same QCD angular momentum can be written in 
an ``interaction-independent'' form\footnote{
Of course, it is not really interaction-free because
the color electric field still depends on the coupling
constant $g$,
$$ 
     \vec E^a = -\vec{\nabla}A^{0a} 
               - {\partial \vec{A} \over \partial t}
               -g f^{abc} \vec{A}^b A^{0c}. $$ 
}  
\begin{eqnarray}
      \vec{J} &=& \int d^3 x  ~\Big[~
          {1\over 2}\bar \psi \vec{\gamma}\gamma_5\psi
          + \psi^\dagger \vec{x}\times (-i\vec{\bigtriangledown})\psi
              \nonumber \\
          &+& \vec{E}\times \vec{A}
          + E_i(\vec{x}\times \vec{\bigtriangledown})A_i ~\Big]\ .
\label{ang1}
\end{eqnarray}    
Because some terms contain explicitly partial derivatives
and gauge potentials, the above expression is not 
{\it manifestly} gauge invariant. 
Nonetheless, the physical meaning of each  term
appears to be  clear. The first term is the quark spin, 
the second  the dynamical (canonical) quark orbital angular
momentum, the third the gluon
``spin'', and the last term the gluon ``orbital'' angular 
momentum.

According to the above decomposition, one
can write down a sum rule for the nucleon spin \cite{jaffemanohar}
\begin{equation}
        {1\over 2} = {1\over 2}\Delta \Sigma(Q^2)
             + L_q'(Q^2) + \Delta g(Q^2) + L_g'(Q^2) \ .
\end{equation}
Here  the matrix elements of the individual
operators are defined in a nucleon state with $p^\mu=(E,0,0,p)$ 
and helicity 1/2, e.g., 
\begin{equation}
     \Delta g(Q^2) = \left \langle p^\mu{1\over 2}\left|\int d^3\vec{x}
(\vec{E}\times\vec{A})^z\right|p^\mu{1\over 2}\right\rangle\ . 
\end{equation}
The $Q^2$ dependence results from the renormalization
of the composite operators. One expects that $L_q'$, $\Delta g$ 
and $L_g'$ are gauge as well as frame dependent. 
The purpose of this paper is to study 
how  the gauge dependence affects 
the physical significance of the individual terms in the above sum rule. 
In   the following we will work in the infinite
momentum frame in which  
the angular momentum operators are defined from the
angular momentum density  $\int d^3 x M^{+ij}$.
In particular, the color electric field $E^i$ is  now $F^{i+}$. 

In a recent paper by Chen and Wang \cite{chen}, it was claimed 
that although the individual operators in Eq. (\ref{ang1})
are gauge invariant, they  have gauge-independent 
matrix elements in the nucleon helicity eigenstates. 
In other words, $L_q'$, $\Delta g$, and $L_g'$
were said to be gauge invariant.
The main point in Ref. \cite{chen} is that the variations
of these operators under a gauge 
transformation may be written as a  commutator
between the QCD angular momentum operator and  some 
other operators,  which has vanishing matrix element 
in a nucleon helicity state. If correct, the theorem 
would have warranted a fresh look at the physical significance  of
$L_q'$, $\Delta g$, and $L_g'$.

We find that the  theorem  is contradicted 
by the following explicit calculation. Consider 
an ``on-shell'' quark in the state of momentum $p^\mu$ and
helicity 1/2. We calculate
the matrix element of the gluon spin operator $S_g^z
=\int d^3x (\vec{E}\times \vec{A})^z$ in perturbation theory.
Choosing the light-cone $A^+=0$, we find at one-loop level
\begin{equation}
     \Delta g = {3\over 2}C_F {\alpha_s\over 2\pi} {\ln \left( 
     Q^2\over \mu^2\right)} \ , 
\end{equation}
where $C_F=(N_c^2-1)/(2N_c)$ with $N_c$ the number of colors, 
$Q^2$ and $\mu^2$ are
the ultraviolet and infrared cutoffs, respectively.
On the other hand, in  the  covariant gauge we have
\begin{equation}
     \Delta g = C_F {\alpha_s\over 2\pi} {\ln \left(Q^2\over 
     \mu^2\right)} \ .
\end{equation}
 A similar discrepancy was 
found upon calculating the matrix element of the same operator
in an ``on-shell'' gluon state. 

The reason for the breakdown of Chen and Wang's 
theorem is a subtle one. In a canonically 
quantized gauge theory, transforming a calculation 
from one gauge to another is not trivial. In 
particular, the Hilbert space in the covariant gauge contains
a nonphysical sector. It is not the same as that 
obtained  after quantizing in an axial gauge. 
For a gauge invariant operator
$\hat O$, the gauge invariance of the matrix element means
\begin{equation}
    \langle P1|O(1)|P1\rangle 
   = \langle P2|O(2)|P2\rangle \ , 
\end{equation}
where $1$ and $2$ label the same physical state 
and operator in different gauges. 
$O(i)$ has the same functional
dependence on the gauge potential $A(i)^\mu$ although
the latter is itself a gauge-dependent operator. 
The transformation from one gauge to another is a ``superunitary'' 
transformation which takes a state in one Hilbert space
to another. The transformation on the gauge potential 
operator is
\begin{equation}
   A(1)^\mu = A(2)^\mu + \partial^\mu \Omega \ , 
\end{equation}
where $\Omega$ is in general a quantum 
operator, not a $c$ number function. 
For instance, in going from the covariant gauge 
to the axial gauge, the gauge tranformation is 
\begin{equation} 
       \Omega = \int^{x^-}d\xi A^+(x^+,\xi,x_\perp) \ , 
\end{equation}
which is a quantum operator because $A^+$ is. 
In Chen and Wang's proof, it was assumed that 
the gauge transformation parameters are $c$ numbers, 
which is a strong restriction. [However, 
in certain special circumstances which we will
not discuss here, the theorem
may be correct in leading order perturbation 
theory.]

Thus the concept of the gluon spin contribution
to the nucleon spin is in general a gauge-dependent 
one. This feature is also reflected in the scale 
evolution of $\Delta g$. 
In the light-cone gauge $A^+=0$, it is well-known 
that $\Delta g$ evolves according to 
the Altarelli-Parisi equation \cite{ap},
\begin{equation}
   {d \Delta g(Q^2) \over d \ln Q^2} 
    = {\alpha_s\over 2\pi} \left( {3\over 2}C_F \Delta
    \Sigma + {\beta_0\over 2}\Delta g(Q^2)\right) \ , 
\end{equation}
where  $\beta_0=11-2n_f/3$ with $n_f$ the number of active 
quark flavors.
In the asymptotic limit $Q^2\rightarrow \infty$, the 
gluon spin grows logarithmically,
\begin{equation}
    \Delta g(Q^2)|_{\rm axial~ gauge} \rightarrow \ln Q^2 , 
\end{equation}
where the coefficient of proportionality is
fixed by nonperturbative physics.

In  the Feynman gauge, the evolution equation
becomes much more complicated.
In fact, the following gauge-variant operators
come to mix with the gluon spin
\begin{eqnarray}
    O_1 &=& -\int d^3x~ {\vec \nabla}A^+_a \times \vec{A}_a \ ,  \\ 
    O_2 &=& -\int d^3x~ gf^{abc} A^{+c}\vec{A}^b\times \vec{A}^a \ .
\end{eqnarray}
[There is no ghost operator here because the ghosts do not
carry spin.] Denote the matrix 
elements of the above operators
in the nucleon helicity states as $a_1$ and $a_2$.   
A lengthy calculation yields the following evolution
equation
\begin{equation}
   {d \over d\ln Q^2} \left(\begin{array}{r}
      \Delta g \\ a_1 \\ a_2 \\ \Delta \Sigma \end{array}
     \right)
    = {\alpha_s\over 2\pi} \left(
      \begin{array}{cccc}
      {7\over 8}C_F-{n_f\over 3} & {5\over 12} C_A & {1\over 12}C_A &  C_F \\
      {23\over 24}C_A & {17\over 12}C_A -{n_f\over 3}  & -{1\over 12}C_A 
       & {1\over 2}C_F \\
      -{3\over 2}C_A& -{3\over 2}C_A & {17\over 24}C_A-{n_f\over 3} & 0 \\
      0 & 0 & 0 & 0  \end{array} \right)       
\left(\begin{array}{r}
      \Delta g \\ a_1 \\ a_2 \\ \Delta \Sigma \end{array}
\right) \ , 
\end{equation}
where $C_A = N_c$. Thus to evolve the gluon spin to a new perturbative 
scale, one  needs not only   the gluon spin at the starting scale 
but also the matrix elements of $O_1$ and $O_2$. 
To find out the asymptotic behavior 
as $Q^2 \rightarrow \infty$, we diagonalize
the upper $3\times 3$ mixing matrix. The three 
eigenvalues are $\lambda_1 = (11/6)
C_A-n_f/3 = \beta_0/2$, $\lambda_2 = (17/24)C_A
-n_f/3$, and finally $\lambda_3 = (11/24)C_A
-n_f/3$. From these, we  found out  that the
leading asymptotic behavior of the gluon spin in the  Feynman 
gauge is the same as that in the light-cone gauge,  
\begin{equation}
     \Delta g(Q^2)|_{\rm Feynman ~ gauge} \rightarrow \ln Q^2 \ . 
\end{equation}
Of course, the coefficients of 
proportionality in the two gauges are different. 

Given that the gluon spin is a gauge-dependent 
concept, it is remarkable that its value 
in the light-cone gauge can be extracted from 
the gluon polarization distribution measurable
in high-energy scattering. What one extracts 
in those experiments is of course gauge-invariant
and is in fact the matrix element of 
the following 
gauge-invariant non-local operator\cite{manohar},
\begin{equation}
    O_g = \int^{\infty}_{-\infty} dx
         {n^-\over 2}\int^{\infty}_{-\infty}
        {d\lambda \over 2\pi}e^{i\lambda x} 
         F^{+\alpha}(\lambda n) e^{-ig\int^\lambda_0
          dyn\cdot A(yn)} \tilde F_\alpha^{~+}(0) \ .   
\end{equation}
However, the physical interpretation of this operator
is in general not obvious. Interestingly,
in the light-cone gauge $A^+=0$, 
the above operator reduces to 
the gluon spin operator $S_g^z$. 
This relationship says nothing about the gauge
transformation property of the gluon spin;
it merely means that the gluon spin in the axial gauge 
can be obtained from the matrix element of a 
gauge-invariant operator. In other words,
the {\it gauge-invariant extension} of the gluon
spin in light-cone gauge can be measured.
[This situation is similar in spirit
to the following example of length in special relativity.
The proper length of a pencil is clearly frame 
independent. When we say the length of a house in the
frame $v=0.9999c$ is the same as the proper length of the
pencil, we are not saying that the length of the
house is  frame-independent. Rather, we are saying 
that the length of the house in a special frame
can be known from measuring a frame-independent 
quantity.] Note that one can easily find   
gauge-invariant extensions of the gluon spin
in other gauges. But we may not always
find an experimental observable which 
reduces to the gluon spin in these gauges.
As far as the nucleon spin structure is concerned,
however, the gluon spin in  the covariant gauge is as interesting
as  its counterpart in the light-cone gauge.

Finally, we turn to the orbital angular momentum operators
in Eq. (\ref{ang1}). The role of the orbital angular 
momentum in parton splitting processes was first
studied by Ratcliffe \cite{rat}. 
In \cite{ji3}, Tang and two of us worked out the
leading-logarithmic scale dependence 
of the orbital angular momenta 
in the light-cone gauge,  
\begin{equation}
      {d\over d\ln Q^2} \left(\begin{array}{c}
                      L_q' \\
                      L_g'
                    \end{array} \right)
    = {\alpha_s(Q^2)\over 2\pi} \left( \begin{array}{rr}
                       -{4\over 3}C_F & {n_f\over 3}\\
                      {4\over 3}C_F & -{n_f\over 3}
               \end{array} \right)
        \left(\begin{array}{c}
                      L_q' \\      
                      L_g'
                    \end{array} \right) +
   {\alpha_s(Q^2)\over 2\pi} \left( \begin{array}{rr}
                       -{2\over 3}C_F & {n_f\over 3}\\
                       -{5\over 6}C_F & -{11\over 2}
               \end{array} \right)
         \left(\begin{array}{c}
                      \Delta \Sigma \\
                          \Delta g
                    \end{array} \right) \ .
\label{hjt}
\end{equation}
The first term on the right-hand side
exhibits the effects of self-generation of
the orbital angular momenta. The second 
represents the generation of orbital angular
momenta from the quark and gluon spin.
The above equation leads to some interesting 
results about the spin structure of the
nucleon in the asymptotic limit. As we are going
to show below, however, the actual operator mixing 
is more complicated than what is shown in the above 
equations although the result in the aymptotic 
limit remains intact.
 
We note that in general there is an additional 
operator which mixes with the quark and gluon 
orbital angular momentum operators,
\begin{equation}
    \Delta L = \int d^3x \psi^\dagger (\vec{x}\times
    (-g\vec{A}))^z\psi \ . 
\end{equation}
Therefore, we proceed to calculate the matrix element
of $\int d^3x \psi^\dagger(\vec{x}\times {(-i)\vec{\nabla}})^z\psi$
in an ``on-shell'' quark-gluon-quark state. 
At the leading-logarithmic order, it contains the
following scale-dependent term
\begin{equation}
     {\alpha_s\over 2\pi}\ln Q^2
    \bar u(x_2p)\left(\not\! n g^{\sigma\perp}
    + {1\over x_1-x_2}\ln (x_2/x_1)
     \left(x_1\not\! n \gamma^\perp\gamma^\sigma
    + x_2\not\! n \gamma^\sigma\gamma^\perp\right)\right)u(x_1p)
   \epsilon_\sigma^*\ .
\end{equation}
This result indicates that the operator that 
mixes with $L_q'$ and $L_g'$
is in fact more complicated than the simple guess
$\Delta L$.  The most general
form is the following non-local operator
\begin{equation}
    \int d^3x \vec{x} \times f(n\cdot\partial_\psi, 
n\cdot\partial_{\psi^\dagger})
        \psi^\dagger \vec{\gamma}\not\!\! A \not\! n\psi 
    + {\rm h. c.} 
\end{equation}
where $\partial_\psi$ and $\partial_{\psi^\dagger}$
are derivatives acting on $\psi$ and $\psi^\dagger$
respectively, and $f(x, y)$ is a function
$x$ and $y$ and takes different forms at different
orders of perturbation theory. Therefore,
we conclude that to evolve the matrix elements of
the gauge-variant orbital angular momentum 
operators is extremely complicated in the light-cone
gauge.\footnote{Note that the light-cone gauge
calculations must be supplemented with some prescriptions
for the light-cone singularities (additional
gauge fixing). In our calculation, we have 
used a prescription such that the regularization
is independent of the minus component of the momenta
flowing through the gluon propagators. In other regularizations,
such as the Mandelstam-Leibbrandt prescription, 
the result can be different \cite{bass}. Of course, 
for studying ing truly gauge-invariant quantities, all prescriptions
are equivalent.} The same statement applies to the 
orbital angular momentum distributions 
defined in Refs. \cite{hs,hk}.

The evolution in the Feynman gauge is again 
different. Here we do not have the problem 
of mixing with infinitely many operators. Apart from
the quark and gluon orbital angular momentum
operators and $\Delta L$, the ghost field also 
carries the orbital angular momentum $L_\omega$.
Thus, a complete evolution equation  
will contain at least the mixing of $L_q'$, $L_g'$, 
$L_\omega$, $\Delta L$ among themselves 
and with $\Delta g$, $\Delta \Sigma$, $a_1$ and $a_2$. 
Because of its limited use, we have not 
calculated the full mixing matrix. However,
we did perform a few calculations just to
explore some of differences. 
We find that the first entry in the evolution
matrix in Eq. (\ref{hjt}) changes from 
$-(4/3)C_F$ in the light-cone gauge to $-C_F/3$
in the Feynman gauge. The evolution of $L_q'$
does depend on $\Delta L$
\begin{equation}
     {dL_q'\over dt} = {\alpha_s\over 2\pi}\left[ \left(-{1\over 3}C_F + {1\over 8}C_A\right)
            \Delta L -{1\over3}C_F L_q'+...\right] \ .
\end{equation}
Conversely, the evolution of $\Delta L$ also 
depends on the other matrix elements
\begin{equation}
     {d\Delta L\over dt} = {\alpha_s\over 2\pi}
\left[-\left(C_F + {1\over 8}C_A\right)
    \Delta L -C_F L_q' +... \right]\ . 
\end{equation}
These equations would be interesting only if
we could find ways to calculate these nonperturbative
matrix elements in the Feynman gauge. 

If the evolution of the gauge-dependent
orbital angular momentum is complicated, how
about their experimental measurement? Is it
possible, for instance, to have a gauge-invariant
extension of the quark orbital angular momentum 
measurable in high-energy scattering 
similar to the gluon spin? A
gauge-invariant operator that reduces 
to the quark orbital angular momentum 
in the light-cone gauge has been discussed 
recently in Ref. \cite{jaf}.
We note, however, that 
non-local operators with dependence
on spatial coordinates have not been seen
in factorization of hard forward 
scattering processes. In particular, 
inclusive deep-inelastic scattering 
does not depend on these types of operators.

Given the difficulty of evolving and measuring
gauge-dependent orbital angular momenta, 
a question arises naturally as how
to incorporate the polarized gluon 
distribution $\Delta g(x)$ in 
unravelling the spin structure of the nucleon, 
particularly since several 
experiments have been proposed to 
measure $\Delta g(x)$ in high-energy processes.
A satisfactory solution can be found by following
the approach outlined in Ref. \cite{hjw} and taking 
seriously the suggestion in Ref. \cite{ji1}.

{}From the off-forward gluon distributions defined from the 
twist-two gluon operators, one can 
introduce the gluon angular momentum distribution \cite{hjw} 
\begin{equation}
    J_g (x) = {1\over 2} x(g(x) + E_g(x)) \ , 
\end{equation}
where $g(x)$ is the unpolarized gluon distribution
and $E_g(x)$ is the forward limit of an off-forward
gluon distribution \cite{ji5}. $J_g(x)$ is gauge invariant, 
evolves like the twist-two gluon distribution, 
and is accessible experimentally. {}From this 
and the gluon helicity distribution $\Delta g(x)$, we
can define the gluon orbital angular momentum 
distribution 
\begin{equation}
    L_g(x) = J_g(x) - \Delta g(x) \ . 
\end{equation}
$L_g(x)$ is experimentally measurable because 
$J_g(x)$ and $\Delta g(x)$ are. 
The evolution equation for $L_g(x)$ is straightforward 
\begin{eqnarray}
{d \over d\ln Q^2}L_{gn}
  &= &\gamma_{gg}(n+1)L_{gn}
    + \gamma_{gq}(n+1)L_{qn} \nonumber \\
  && + \left(\gamma_{gg}(n+1)-\Delta \gamma_{gg}(n)\right)\Delta g_n
    + \left({1\over 2}\gamma_{gq}(n+1)
    -\Delta \gamma_{gg}(n)\right)\Delta \Sigma_n \ , 
\end{eqnarray}
where $\gamma_{ij}$ and $\Delta \gamma_{ij}$ 
are the anomalous dimensions for the 
spin-independent and spin-dependent 
twist-two operators \cite{ap}. 
However, the catch here is that 
$L_g(x)$ can be interpreted
as the gluon orbital angular momentum distribution
only in the light-cone gauge. If one studies the gluon 
orbital angular momentum, say in a covariant gauge,
$L_g(x)$ would not be sufficient.

\acknowledgements
PH thanks the Fulbright Foundation for sponsoring 
his visit to the University of Maryland. 
This work is supported in part by funds provided by the
U.S.  Department of Energy (D.O.E.) under cooperative agreement
DOE-FG02-93ER-40762.


\begin{references}
\frenchspacing

\bibitem{review}
See for example, H. Y. Cheng, Int. J. Mod. Phys. A {\bf 11}, 5109 
(1996). 

\bibitem{ji1}
X. Ji, Phys. Rev. Lett. {\bf 78}, 610 (1997). 

\bibitem{fey}
R. P. Feynman, {\it The Feynman lectures on physics}, 
Vol. III, Addison-Wesley, Reading, MA, 1965. 

\bibitem{ji2}
I. Balitsky and X. Ji, Phys. Rev. Lett. {\bf 79}, 1225 (1997).
 
\bibitem{bro}
V. Barone, T. Calarco, and A. Drago, hep-ph/9801281. 

\bibitem{sin}
D. Singleton and V. Dzhunushalev, hep-ph/9807239.

\bibitem{jaffemanohar}
R. L. Jaffe and A. Manohar, Nucl. Phys. {\bf B337}, 509 (1990).

\bibitem{chen}
X. Chen and F. Wang, hep-ph/9802346.

\bibitem{ap}
G. Altarelli  and G. Paris, Nucl. Phys. {\bf B126}, 298 (1977). 
 
\bibitem{manohar}
A. Manohar, Phys. Rev. Lett. {\bf 66}, 289 (1991). 
 
\bibitem{rat}
P. G. Ratcliffe, Phys. Lett. {\bf B192}, 180 (1987). 

\bibitem{ji3}
X. Ji, J. Tang and P. Hoodbhoy, Phys. Rev. Lett. {\bf 76}, 740 (1996). 

\bibitem{bass}
A. Bassetto, hep-ph/9605421, Invited talk at 
the Workshop ``QCD and QED in Higher Order'', Rheomsberg, 
April, 1996.  

\bibitem{hs}
P. H\"agler and A. Sch\"afer, hep-ph/9802362.

\bibitem{hk}
A. Harindranath and R. Kundu, hep-ph/9802406.

\bibitem{jaf}
S. V. Bashinskii and R.L. Jaffe, hep-ph/9804397.

\bibitem{hjw}
P. Hoodbhoy, X. Ji and W. Lu, hep-ph/9804337.

\bibitem{ji5}
X. Ji, Phys. Rev. {\bf D55}, 7114 (1997).

\nonfrenchspacing
\end{references}
\end{document}